# Rapid classification of quantum sources enabled by machine learning

Zhaxylyk A. Kudyshev[1,2]†, Simeon Bogdanov[1]†, Theodor Isacsson[1,3], Alexander V. Kildishev[1]

Alexandra Boltasseva[1*], and Vladimir M. Shalaev[1]

[1]School of Electrical and Computer Engineering, Birck Nanotechnology Center and Purdue Quantum Science and Engineering Institute, Purdue University, West Lafayette, IN, 47906, USA

[2]Center for Science of Information, Purdue University, West Lafayette, IN, 47906, USA

[3]Department of Applied Physics, KTH Royal Institute of Technology, Stockholm, Sweden

**Abstract:** Deterministic nanoassembly may enable unique integrated on-chip quantum photonic devices. Such integration requires a careful large-scale selection of nanoscale building blocks such as solid-state single-photon emitters by the means of optical characterization. Second-order autocorrelation is a cornerstone measurement that is particularly time-consuming to realize on a large scale. We have implemented supervised machine learning-based classification of quantum emitters as "single" or "not-single" based on their sparse autocorrelation data. Our method yields a classification accuracy of over 90% within an integration time of less than a second, realizing roughly a hundredfold speedup compared to the conventional, Levenberg-Marquardt approach. We anticipate that machine learning-based classification will provide a unique route to enable rapid and scalable assembly of quantum nanophotonic devices and can be directly extended to other quantum optical measurements.

†authors with equal contributions; *corresponding author aeb@purdue.edu

**One Sentence Summary:** Sub-second characterization of the quantum emitters

Integrated quantum photonics has recently emerged as one of the key enablers for the quantum information science and technology (QIST)(*1*) . Typically, quantum photonic circuits are realized using nonlinear sources of single photons that operate probabilistically and do not allow the generation of large multi-photon states(*2*). As a result, the number of photonic qubits in quantum circuits is currently limited to twenty(*3*). Alternatively, solid-state quantum emitters(*4*) have recently reached near-ideal single-photon characteristics(*5*). Successful implementation of quantum photonic circuits based on quantum emitters depends crucially on the selection of these emitters from a large inhomogeneous set. It requires efficient identification of bright, stable single-photon emitters with fast emission rates, high quantum yield, and narrow optical linewidth. Since a universal platform for integrated quantum photonics is lacking, hybrid integration methods are being actively explored (*6*). Recently, deterministic atomic force microscopy-assisted assembly has been utilized to realize hybrid nano-structures for ultrafast emission of single-photons at room temperature(*7*). As a rule, micromanipulation methods are needed to integrate devices across different material platforms(*8*). With the growing interest in scalable realization and rapid prototyping of quantum devices, high-speed, robust classification of "single" or "not-single" emitters becomes of utmost significance(*9–14*).

For optical quantum emitter characterization, photon anti-bunching measurements using second-order auto-correlation have long been used to measure the single-photon purity of the emission. Such experiments are typically implemented using a Hanbury-Brown-Twiss (HBT) interferometer, composed of a beamsplitter directing the emitted light to two single-photon detectors connected to a correlation board (Fig. 1a)(*15*). The correlation board collects pairs of clicks generated by different detectors (detector 1, then detector 2 or vice versa) and bins these co-detection events according to the time delay $\tau$ between the detectors' clicks. Depending on which detector clicks first, the delay is either considered positive or negative. The distribution of co-detections as a function of $\tau$ is described by the second-order autocorrelation function of the emission $g^{(2)}(\tau)$. This distribution is normalized so that $g^{(2)}(\infty)=1$, indicating that there is no correlation between the detection events spaced by an infinite time interval. In practice, the number of co-detection events collected with this method tends to 0 as $\tau$ increases. Therefore, the maximum $\tau$ in the measured distribution is kept at a finite value that is much smaller than the inverse photon detection rate, but much larger than the time constants of the emitter photodynamics.

Commonly, the metric for the single-photon purity of the emission is the value of $g^{(2)}(0)$, indicating the normalized number of coincidental detections(16). The measurement of single-photon purity is of paramount importance for characterizing quantum emitters. Besides describing the quality of quantum emission, it contains information about the fluorescence lifetime and can be used to correct other data such as optical saturation curves(17) and photon indistinguishability(18). Fig. 1b shows examples of autocorrelation datasets with different acquisition times. In practice, the co-detection rate is proportional to the square of source intensity, which is very low for single-photon emitters. At the same time, the conventional statistical classification with the Levenberg-Marquardt (L-M) fit requires a sufficiently populated histogram to reliably retrieve the value of $g^{(2)}(0)$ (see supplementary materials, methods section). As a result, determining $g^{(2)}(0)$ with sufficient accuracy is highly time-consuming, thus hindering the development of scalable techniques for photonic QIST device assembly and prototyping. Faced with the challenge of rapidly identify emitters with high single-photon purity from a large set, one may however define a heuristic threshold for the value of $g^{(2)}(0)$ and only strive to fully characterize the emitters that meet this threshold.

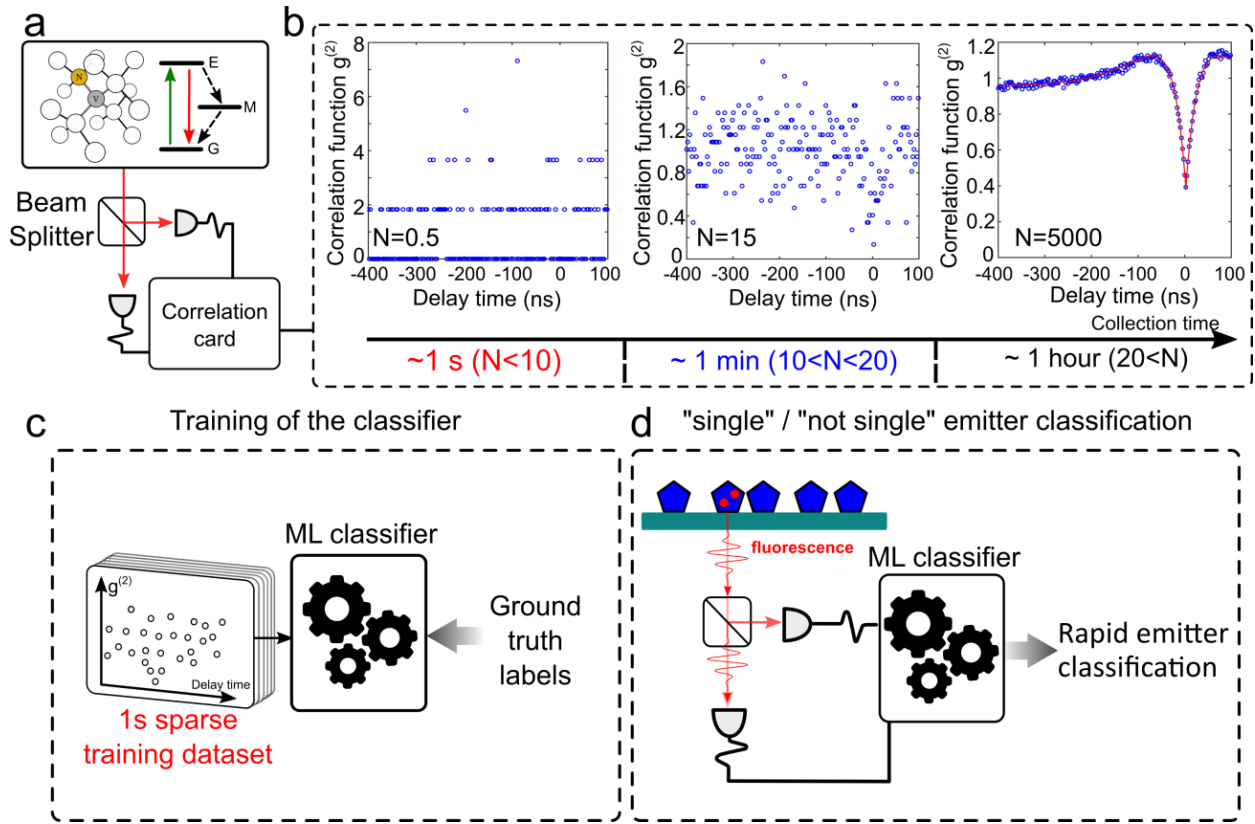

**Fig. 1. Rapid binary classification of quantum emitters enabled by machine learning.** (a) Fluorescence from diamond nitrogen vacancy (NV) sources (single centers and center ensembles) is analyzed by Hanbury-Brown-Twiss (HBT) autocorrelator. The emitters are modeled as three-level quantum optical systems. (b) Examples of autocorrelation datasets with different acquisition times: 1s, 1min, and 1hour. The Levenberg-Marquardt (L-M) fitted complete datasets yield the actual values of $g^{(2)}(0)$ used for training and assessment of the classification accuracy. (c) Classifier network training. Sparse data collected from HBT measurement are used as a training set. The $g^{(2)}(0)$ values retrieved from L-M fit of a complete dataset from each corresponding emission source are supplied to the classifier in training as the ground truth. (d) Machine learning-assisted rapid single photon emitter detection. The trained model classifies the previously unencountered emission sources as "single" and "not-single" emitters based on sparse autocorrelation data.

Theoretically, the autocorrelation at zero delay from an ensemble of $n$ identical, equally contributing emitters is equal to $1-n^{-1}$, yielding $g^{(2)}(0)=0$ for a single emitter(*19*). In practice, single-photon purity is always finite due to experimental imperfections, such as background radiation and detector dark counts, and thus the measured $g^{(2)}(0)$ is always strictly positive. Moreover, emitters may not be identical or equally contributing. These uncertainties may lead to any value of $g^{(2)}(0)$ between 0 and 1, thus further blurring the boundary between the "single-photon" and "classical" emission regimes. These regimes are still heuristically defined by $0<g^{(2)}(0)<0.5$ and $g^{(2)}(0)\geq 0.5$ respectively(*20*). In the rest of the paper, we consider 0.5 as our classification threshold, although a different value may be chosen in practice (see supplementary materials). Despite its clear necessity, high-speed, accurate emitter classification based on single-photon purity has so far remained elusive.

In the last decades, different methods of machine learning (ML) have attracted significant interest in the optics community(*21–26*). Recently ML algorithms have been applied to quantum optical problems(*27–30*). Combining the Bayesian phase estimation with Hamiltonian Learning techniques for analyzing large datasets from nitrogen vacancy (NV) centers in bulk diamond allowed magnetic field measurements with extreme sensitivity at room temperature(*31*). Hamiltonian Learning was adopted for the characterization of different quantum systems(*32*), including the characterization of electron spin states in diamond NV centers(*33*). The development of autonomous adaptive feedback schemes allows incorporating decision mechanisms into quantum measurements(*34, 35*). Yet, one of the most powerful applications of ML algorithms relates to object classification problems that encompass most quantum optics measurements, including emitter classification. Machine learning can dramatically speed up quantum measurements, thus transforming the area of quantum photonic testing, assembly, and prototyping.

In this work, we implemented supervised ML algorithms to classify single vs not-single photon emitters using the sparse autocorrelation data. We have applied four different supervised ML classifiers (support vector classification (SVC) (*36*), gradient boosting classifier (GBC) (*37*), voting classifier (VC) (*38*) and convolutional neural network (CNN) (*39*)) to map autocorrelation measurement data into "single"/"not-single" categories. First, autocorrelation data was acquired on a set of emitters with an integration time of several minutes, allowing an accurate retrieval of $g^{(2)}(0)$ using the **st**andard L-M fitting. For these datasets that we refer to as "complete", the L-M fitted $g^{(2)}(0)$ values yield an error of about 0.03 and can be regarded as the ground truth. These complete datasets were further sectioned into "sparse datasets" with integration time of 1s. The classifier was trained (Fig. 1c) on sparse datasets. Subsequently, sparse datasets from previously unencoutered NVs were fed into the trained classifier, and the classification accuracy was statistically assessed (Fig.1d). Both the training of the ML classifier and the estimation of classification accuracy were based on the ground truth $g^{(2)}(0)$ values retrieved from the L-M fits of the complete datasets. We show that ML-assisted classification remains highly accurate even for sparse data, whereas the conventional L-M fit applied to the same sparse data fails and performs no better than a random guess.

**Results**

**Theoretical framework**

To determine the best ML-assisted algorithm for the sparse data classification, we first performed a numerical experiment. In the numerical experiment, the emulated autocorrelation data was fed into different types of ML algorithms targeting their classification based on the $g^{(2)}(0)$ threshold value of 0.5. The autocorrelation data were obtained by emulating results of an HBT experiment with photophysical emitter parameters similar to those featured by the NV centers under investigation. Each emitter was modeled as a three-level system with an excited state (E) radiatively coupled to a ground state (G), and a metastable state (M) non-radiatively coupled to both E and G (Fig. 1a). We simulated the autocorrelation experiment by counting co-detection events from two virtual "detectors" and binning these events according to the time delay between the two "detector" clicks in each co-detection event. An elementary numerical experiment was repeated, until obtaining the desired number of such events (see supplementary materials, methods section). To generate autocorrelation histograms with a range of $g^{(2)}(0)$ values, we assumed that the emission was produced partly by a single quantum emitter and partly by photonic background presenting no autocorrelation features. To emulate the emission with a certain input level of antibunching $g^{(2)}(0)$, we set the intensity fraction of the photonic background in the total emission to be $r_{bg} = 1 - \sqrt{1 - g^{(2)}(0)}$. More details on the numerical experiment and the underlying model are provided in supplementary text S1.

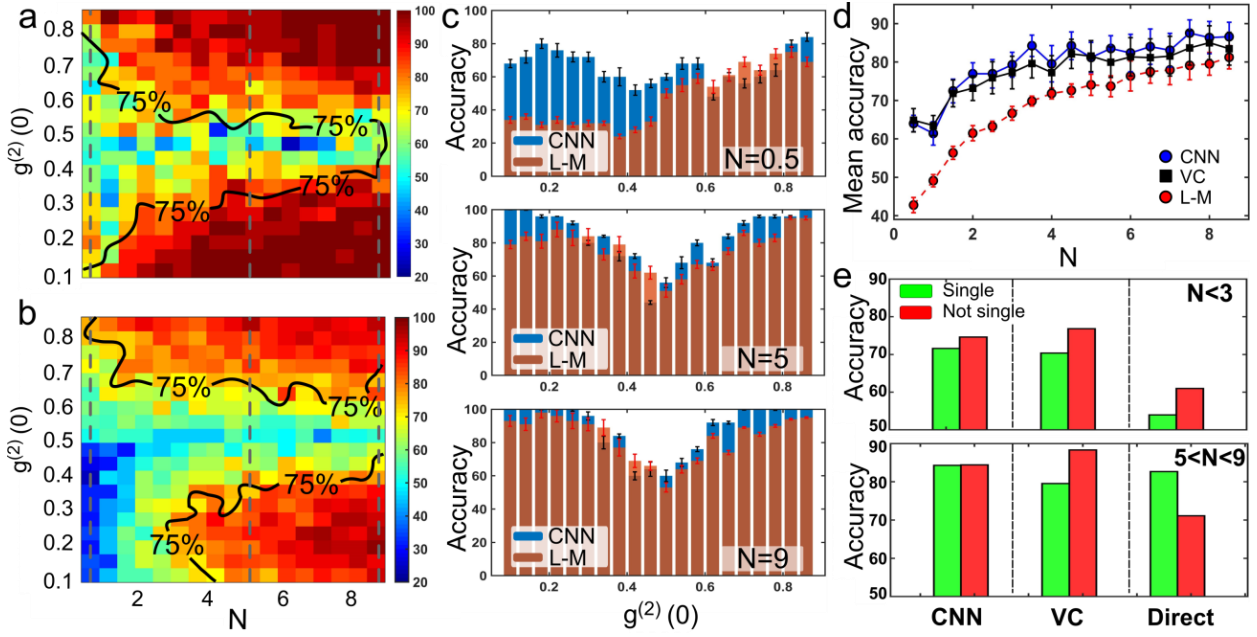

**Fig. 2. Classification based on ETCE model data set.** (a) Accuracy distribution of the CNN classifier, as a function of $g^{(2)}(0)$ and $N$. (b) Accuracy distribution of the direct method, done by L-M fitting. The contour outlines the boundary of >75% region; (c) Comparison between the direct fitting (brown histograms) and CNN-based (blue histograms) classifications, for different average count per bin cases: $N = 0.5$ (top), $N = 5$ (middle), $N = 9$ (bottom). Corresponding cross section positions are shown with dashed, vertical lines in sections (a) and (b). (d) Averaged accuracy over all $g^{(2)}(0)$ realizations as a function of luminescence level $N$: CNN classifier (blue circles), voting classifier (open squares), L-M fitting (red circles). (e) Accuracy distribution for two classes for low ($N < 3$) and moderate emission regions ($5 < N < 9$). Error bars show the corresponding standard error of the mean values.

The theoretical model included two variable parameters: the ground truth value of $g^{(2)}(0)$, and the average co-detection counts per bin $N$. The latter parameter was defined as the total number of co-detection events in the dataset divided by $N_{bins}$. We generated forty thousand datasets, with $g^{(2)}(0)$ spanning the interval between 0.1 and 0.9 with a step of 0.04 and $N$ spanning from 0.5 to 10 with a step of 0.5. One hundred datasets were generated for each combination of $g^{(2)}(0)$ and $N$. 70% of all the datasets were used for training, and the remaining 30% served to test performance of the classifiers. The accuracy of "single" vs. "not-single" emitter classification is defined as $F = N^{(correct)} / N^{(tot)}$, where $N^{(correct)}$ is a number of correct guesses, and $N^{(tot)}$ - total number of realizations for a given combination of $g^{(2)}(0)$ and $N$.

We applied 4 different supervised machine learning classification techniques to the generated datasets: (i) SVC, (ii) GBC, (iii) VC (a combination of logistic regression(*40*) and k-nearest neighbors (k-NN) algorithm(*41*)) and (iv) CNN based binary classifier. The performance and detailed description of each model is provided in the supplementary text S2. Here, we outline the operation and results of the two classification algorithms, CNN and voting classifier (VC), that showed the best performance.

The CNN binary classifier consists of one input layer, three hidden convolutional layers, one max-pooling layer followed by two fully connected layers. The input layer has a dimension of $N_{bins}$, the same as the dimension of the autocorrelation dataset. The three hidden layers of the CNN classifier are extracting the main features of this dataset, while the preceding two fully connected layers perform binary classification of the dataset based on these main features. The training process was performed with the stochastic gradient descent optimization of the weight using the dataset and corresponding ground truth values ("single" or "not-single" based on $g^{(2)}(0)$ values). More details on the CNN performance and "decision making" process can be found in the supplementary materials (section S5).

Along with the CNN, we adapted the VC method built on logistic regression (LG) and k-NN algorithm. The main idea of the VC is to average the results of several pre-trained different classifiers. The voting classification is realized in two steps. First, LG and k-NN classifiers are independently trained on the same ensemble of autocorrelation datasets. Then, both trained models are applied to the test ensemble of datasets. The voting classification is obtained by weighted averaging of the two outputs with 2 to 1 weight ratio between the LG and k-NN classifiers.

Figure 2a shows the accuracy map of the CNN-based classifier plotted as a function of $g^{(2)}(0)$ and $N$. For comparison, we also plotted the classification accuracy of the L-M fit for the same ensemble of datasets (Fig. 2b). The black curve shows the contour line corresponding to 75% accuracy. The CNN-based classifier enables over 75% accuracy within a larger parameter space area, whereas the L-M fit (i) performs poorly for the datasets with small $N$ and (ii) breaks down completely in the region of $N < 3$. The asymmetry of the L-M fitting accuracy in this region comes from the large uncertainty in the normalization of sparse datasets which skews the fit towards larger values. The classification performance of the L-M fit and CNN-based method is further analyzed in Fig. 2c. The figure presents the cross-sections of the accuracy maps of Fig. 2a and Fig. 2b for constant values of $N = 0.5$, 5 and 9. In nearly all the cases, CNN outperforms L-M fitting.

For $N = 0.5$, L-M fitting leads to unbalanced accuracy distribution between "single" and "not-single" classes resulting in <50% mean accuracy, while the CNN classifier ensures 70% accuracy in this region. A trough in accuracy consistently occurs close to the decision boundary between two classes, corresponding to $g^{(2)}(0) = 0.5$. With increasing $N$, the accuracy trough becomes expectedly narrower, and in the limit of high $N$ both techniques (CNN and L-M) asymptotically arrive at 100% classification accuracy (supplementary materials, section S2). Figure 2d compares the classification accuracy of L-M fit, CNN and VC methods, averaged over all of the $g^{(2)}(0)$ values as a function of $N$. The CNN-based classifier strongly overperforms the L-M fit and features slightly better performance in comparison with the VC method.

Another important aspect of the classification is accuracy distribution between classes since imbalanced classification leads to biased predictions and misleading classification accuracies. Figure 2e shows the accuracy distribution between "single" and "not-single" classes for the CNN-based classifier, VC method, and L-M fits. The CNN-based classifier exhibits a more balanced accuracy distribution between "single" and "not-single" classes in comparison with the VC method, while the L-M fitting shows imbalanced performance for both sparse data region (N < 3) and region with 5 < N < 9.

For $N < 3$ the L-M fit has a low mean accuracy of $57\% \pm 3.7\%$ ($53\% \pm 5\%$ for "single", $61\% \pm 1.32\%$ for "not-single" emitters). The CNN classifier has the $\sim 73\% \pm 3\%$ mean accuracy rate ($71\% \pm 3\%$ for "single", $75\% \pm 3.1\%$ for "not-single" emitters), while the VC has a $74\% \pm 2.4\%$ ($70\% \pm 2\%$ and $76.8\% \pm 2.87\%$ for "single"/"not-single" emitters respectively). Along with the accuracy score additional performance matrices of classification, like precision, recall, F1-score and confusion matrices are analyzed (supplementary materials, section S4). Our analysis clearly shows that on a broad range of emitter parameters, ML-based approaches perform significantly better than the L-M fit. The performance difference is especially striking for the sparsest datasets, which is particularly important for speeding up the single-photon emitter characterization.

**Experimental emitter classification**

Having selected the best performing network architectures, we test both the CNN and VC classifiers for emission classification from physical nanodiamond-based NV centers. In the experiment, we used nanodiamonds (Fig.3a inset) randomly dispersed on a coverslip glass substrate with at least some of them containing single or multiple NV centers. Two avalanche detectors (D1, D2) featuring about 30 ps jitter were used for single-photon autocorrelation measurements. Time-correlated photon counting was performed by a correlation card with a 4 ps internal jitter. The total histogram span was set to 500 ns, and the co-detection events were collected into 215 equally sized time bins. The first 171 bins collected co-detections with a negative delay (D2 clicks first), while the remaining 43 bins corresponded to a positive delay (D1 clicks first). Thus the maximum absolute observable delay in a co-detection event was $t_{max} \approx 400$ ns. Figure 3a summarizes the experimental setup.

Photon autocorrelation measurements were performed on a set of 41 emitters. For each emitter, the sparse datasets were compounded into a complete dataset, from which the ground truth $g^{(2)}(0)$ value was attained using the L-M fitting algorithm. These fitted values of $g^{(2)}(0)$ on full datasets were appended to each corresponding sparse dataset as a label and used as the ground truth for the training/testing purposes. The L-M fitting uncertainty for the complete datasets varied from ±0.01 to ±0.05, with the average of ±0.03.

In total, 9416 sparse physical datasets were collected from 15 "single" and 26 "not-single" emitters (as determined by L-M fit of their full datasets). The numbers of datasets acquired for each emitter are represented as circle sizes in Fig. 3b and plotted as a function of their $N$ and $g^{(2)}(0)$. Most of the experimental datasets exhibit $N < 5$ and fall in the range where a significant improvement in classification accuracy is expected from ML-assisted methods. More details regarding the experimental datasets (dataset size, ground truth values, the fitting errors for each emitters) are given in the supplementary materials (section S7).

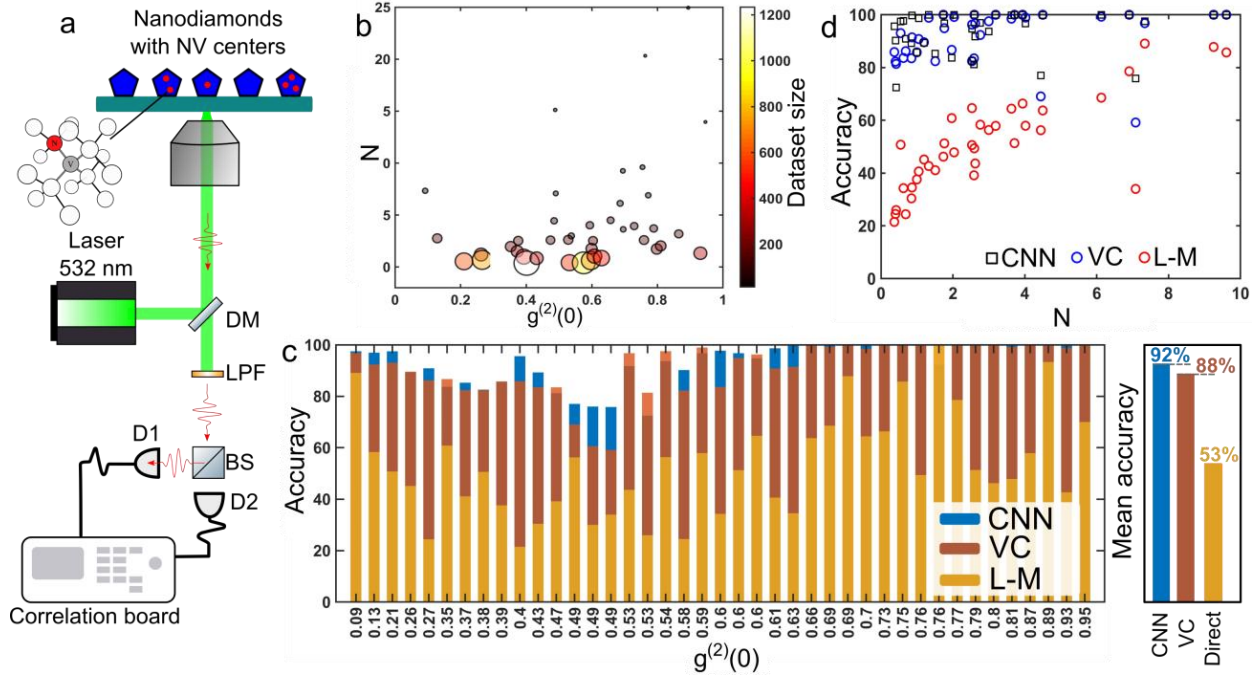

**Fig. 3. Classification of physical autocorrelation data.** (a) Schematics of the HBT interferometer. Labels: DM – dichroic mirror; LPF – long-pass filter; BS – beam splitter; D1/D2 – detectors. (b) Distribution of datasets as a function of $g^{(2)}(0)$ and $N$. Each circle corresponds to an ensemble of datasets for a given emitter and therefore it shares the same values of $g^{(2)}(0)$ and $N$. The circle radius is proportional to the number of sparse datasets acquired for the corresponding emitter. (c) Comparison of classification accuracy by the CNN(blue bars), VC (brown bars) and L-M fitting (yellow bars) methods for each emitter. $g^{(2)}(0)$ for each emitter are displayed on the horizontal axis. Bar plot on the right shows mean accuracy of three methods averaged over all emitters classification. (d) Accuracy as a function of average count per bin values of CNN (black squares), voting classifier (blue markers) and L-M fitting (red markers).

The analysis of dataset size influence on the performance of CNN and VC classifiers reveals that the CNN based approach requires a larger amount of training data in comparison with VC method (supplementary materials, section S3). Moreover, for balanced training of the VC and CNN classifiers, it is necessary to have an equal amount of datasets for both classes. Since the "single" emitter class has smaller amout of dataset, the datasets for this class were upsampled to 30% using the bootstrapping method during the training phase.

To assess the classification accuracy of a given emitter, we trained the network on all the datasets, except for those belonging to that given emitter. The datasets corresponding to the emitter of interest were then used for classification testing. That approach provided significant data augmentation and allowed for a better training of the classifiers.

Figure 3c shows the accuracy comparison between the CNN, VC and L-M fitting methods as a function of $g^{(2)}(0)$. For all the emitters CNN and VC classifiers expectedly outperform the L-M fitting approach. Noting that the drop in the accuracy of CNN and VC classification happens around the decision boundary region $g^{(2)}(0) \approx 0.5$. However, the accuracy trough is not as pronounced as expected from numerical experiment. This can be explained by the fact that emitters with $g^{(2)}(0)$ close to 0.5 have relatively high emission rates ($N > 4.4$) (supplementary materials, section S7), which allows a more accurate classification. For the "single" ("not-single") emitters the CNN classifier ensures $87\% \pm 1.99\%$ ($97\% \pm 1.17\%$), VC approach yields $82.46\% \pm 2.86\%$ ($96.50\% \pm 1.15\%$) accuracy, while the L-M fit only attains $44.6\% \pm 4.47\%$ ($57.99\% \pm 3.99\%$), which is no better than a random guess within statistical uncertainty. The bar plot on the right shows the mean accuracy of all three methods averaged over all emitters. Additional performance metricses of the CNN and VC classifiers are provided in supplementary materials, section S4. Fig. 3d shows the classification accuracy for all the three methods as a function of $N$. For datasets with $N \leq 5$, the CNN classifier ensures 93% accuracy; the VC algorithm features an 91% average accuracy, while L-M fitting features only 46%. For the $5 < N < 10$, these three methods provide 96%, 92% and 74% accuracy values, respectively. Thus, in the physical experiment, the best ML-assisted algorithm features an equally strong classification advantage with respect to the fitting algorithm, fully confirming the results of the numerical experiment.

**Discussion**

Photon correlation measurements are at the heart of many quantum optical experiments. These measurements require collecting statistics over relatively rare co-detection events and therefore are inherently slow. Obtaining the desired information from a limited number of such co-detection events would be critical for characterizing and assembling quantum optical systems paving the way to scalable integrated quantum circuits.

We have implemented machine learning classification methods for rapidly identifying single emitters based on the value of the photon autocorrelation at zero delay. We analyzed sparse autocorrelation data and performed binary statistical classification ("single"/"not-single" emitter) on sparse datasets that cannot be characterized by a conventional fitting method. We demonstrated that by applying ML-assisted classifiers on the sparse datasets it was possible to achieve over 90% accuracy. The fitting method applied to sparse datasets performed no better than a random guess and required a two orders of magnitude longer, collection time to reach an accuracy of 90%.

The proposed approach has the potential to dramatically advance most quantum optical measurements that can be formulated as a binary or multi-class classification . It can also be straightforwardly extended into a multi-bin classification, providing predictive estimates of $g^{(2)}(0)$ faster than any conventional fitting algorithm. In addition, our approach could also transform higher-order autocorrelation measurements because their datasets can be even more sparse for the same acquisition times. As an example application, our technique can speed up super-resolution microscopy based on single-photon autocorrelation measurements(*42*, *43*), that are currently limited by long image acquisition times. For a precise determination of the autocorrelation value at zero delay however, one would need to employ to regressive machine-learning techniques.

Supervised machine learning algorithms could be efficiently applied to characterize other properties of quantum emitters. For example, the NV center's electron and nuclear spin states can be optically read out through spin-dependent fluorescence intensity. Single-shot readout of the electronic spin at room temperature is a highly-sought-after goal which could be achieved using classification ML approaches. In combination with the recently proposed idea of plasmonic cavity-single photon emitter integration, which allows one to significantly increase the photon count rate of the NV centers(*44*, *45*), ML assisted approach can open up a road for room temperature single-shot sensitive spin readout.

The proposed approach could also have a strong impact on single-molecule spectroscopy. The single-molecule optical signal is often unstable and could be lost for extended periods due to long-term shelving in non-radiative states. The use of ML-assisted signal processing can help with acquiring the necessary information from optically unstable emitters before the optical signal is lost.

To conclude, we demonstrated that supervised machine learning can significantly advance quantum measurements allowing for rapid single-photon source detection, which could enable scalable, rapid quantum device testing, assembly and prototying as well as real-time, precision metrology in quantum materials.

# Supplementary Material for
# Rapid classification of quantum sources enabled by machine learning

*Zhaxylyk A. Kudyshev[1,2]†, Simeon Bogdanov[1]†, Theodor Isacsson[1,3], Alexander V. Kildishev[1]*

*Alexandra Boltasseva[1*] and Vladimir M. Shalaev[1]*

[1]*Birck Nanotechnology Center and Purdue Quantum Science and Engineering Institute,*

*Purdue University, West Lafayette, IN, 47906, USA*

[2]*Center for Science of Information, Purdue University, West Lafayette, IN, 47906, USA*

[3]*Department of Applied Physics, KTH Royal Institute of Technology, Stockholm, Sweden*

**METHODS. Monte-Carlo simulation of the autocorrelation experiment.** The NV center was modeled as a three-level system (Figure 1a, main text) consisting of an excited state E, a ground state G and a metastable state M. We fixed the inter-level transition rates as $\gamma_{GE} = \gamma_{EG} = (50 \text{ ns})^{-1}$, $\gamma_{EM} = (10 \text{ ns})^{-1}$, and $\gamma_{MG} = (150 \text{ ns})^{-1}$. The radiative rate is taken to be similar to that in previously studied NV centers in nanodiamonds(*1*), while the non-radiative constants that are not affected by the photonic environment were taken from another study(*2*). The emission is supposed to obey the corresponding time-dependent probability given in the supplementary section S1. We simulated the autocorrelation experiment by counting co-detection events from two virtual "detectors" and binning these events according to the time delay between the two detector clicks in each co-detection event. We set a histogram bin size $\delta t = 2.34$ ns and a maximum absolute delay between detection events $t_{max} = 400$ ns. We then set the average co-detection rate per bin to $R = (20 N_{bins})^{-1}$ to minimize histogram bias due to the natural preference of the experiment to detect early photons. The simulation was conducted in a series of elementary numerical experiments. Each such experiment proceeded as follows. One virtual "detector" received a photon and started the time counter. The photon arrival at a second virtual "detector" was simulated at each elementary time step of duration, $\delta t$. In the case of the second photon arrival, the time counter was stopped and a co-detection event was recorded in the bin corresponding to the current value of the time counter, thus ending the elementary experiment. The non-arrival of the second photon increased the time counter by $\delta t$, and the experiment continued. If the time counter exceeded $t_{max}$, the elementary experiment was stopped, and no co-detection event was recorded. These elementary experiments were repeated until obtaining the desired number of co-detection events.

**Experimental setup.** The sample with NV nanodiamonds was prepared by cleaning the coverslip sample with solvents, treating it with ultraviolet radiation for an hour and drying a 5 µL droplet of a sonicated nanodiamond solution (20 nm average size, Adamas Nano) on the coverslip surface. All the optical characterization was performed using a custom-made scanning confocal microscope with a 50 μm pinhole based on a commercial inverted microscope body (Nikon Ti−U). To locate the emitters, objective scanning was performed using a P-561 piezo stage driven by an E-712 controller (Physik Instrumente). The optical pumping in the CW experiments was administered by a continuous wave 532 nm laser (Shanghai Laser Century). A power on the order of 1 mW (measured before entering the optical objective) was used to pump the NV centers. The excitation beam was reflected off a 550 nm long-pass dichroic mirror (DMLP550L, Thorlabs), and a 550 nm long-pass filter (FEL0550, Thorlabs) was used to filter out the remaining pump power. Two avalanche detectors with a 30 ps time resolution and 35% quantum efficiency at 650 nm (PDM, Micro-Photon Devices) were used for single-photon autocorrelation measurements. Time-correlated photon counting was performed by a "start-stop" acquisition card with a 4 ps internal jitter (SPC-150, Becker & Hickl). The total histogram span was set to 500 ns and the co-detection events were collected into 215 time bins.

Autocorrelation measurements on each emitter were performed by repeating acquisitions over a period of one second, until accumulating about 300 co-detection events per bin in total. The results of individual measurements were summed up to obtain low-noise autocorrelation data. To extract an estimate of the autocorrelation at zero delay, the summed up autocorrelation histogram was fitted according to a three-level emitter model using the Levenberg-Marquardt (L-M) fit by the following function:

$$g^{(2)}(t) = 1 - A_1 e^{-t/t_1} + A_2 e^{-t/t_2},$$

where $A_{1,2}$ and $t_{1,2}$ are fitting parameters.

# SUPPLEMENTARY SECTIONS

## S1. Monte-Carlo simulation of autocorrelation measurement

At the beginning of each elementary experiment, the time counter is started, signifying the arrival of the first photon. The experiment is stopped either when a co-detection event is completed by the arrival of the second photon or when the timer exceeds the maximum absolute delay $t_{max} = 400$ ns. After the start of the time counter, we simulated the arrival of the second photon within a time delay of $\delta t = 2.34$ ns with respect to the first photon (corresponding to the bin #1). The probability of this event is denoted as $P_1$. If the second photon did not arrive within $\delta t$, we then simulated its arrival between $\delta t$ and $2\delta t$ (corresponding to the bin #2) with probability $P_2$, etc. The elementary experiments were repeated until reaching the pre-determined average number of co-detections per bin $N$.

| Model Parameter | Interpretation | Value |
| --- | --- | --- |
| $t_{max}$ | Maxiumum absolute delay | 400 ns |
| $\delta t$ | Histogram bin width | 2.34 ns |
| $N_{bins}$ | Total number of histogram bins | 215 |
| $\gamma_{EG}$ | Radiative excited-state decay rate | 20 MHz |
| $\gamma_{GE}$ | Excitation rate | 20 MHz |
| $\gamma_{EM}$ | Non-radiative shelving rate | 10 MHz(3) |
| $\gamma_{MG}$ | Non-radiative de-shelving rate | 7 MHz(3) |

**Table S1.** Summary of the model parameters used in the simulation of second-order autocorrelation measurements.

To produce autocorrelation histograms with a range of $g^{(2)}(0)$ values, we assumed that the emission was produced partly by a single quantum emitter with an excited state lifetime of

$\tau = \gamma_{EG}^{-1} = 50$ ns subject to a half-saturating excitation rate $\gamma_{GE} = \gamma_{EG}$, and partly, by a photonic background presenting no autocorrelation features. To simulate the emission with a certain level of antibunching $g^{(2)}(0)$, we took the fraction of the photonic background in the total emission to be $r_{bg} = 1 - \sqrt{1 - g^{(2)}(0)}$. We label the probability of the second photon arrival into the $n$-th bin as

$$P_n = P\left(\begin{array}{c}2^{nd}\text{ photon arrives between} \\ (n-1)\delta t \text{ and } n\delta t\end{array} \middle| \begin{array}{c}2^{nd}\text{ has not arrived between} \\ 0 \text{ and } (n-1)\delta t\end{array}\right).$$

For each bin, that probability was given by:

$$P_n = Rr_{bg} + R(1-r_{bg})\left[1 - (1+a)\exp\left(-\left|n - n_0 + \frac{1}{2}\right|\delta t \lambda_1\right) + a\exp\left(-\left|n - n_0 + \frac{1}{2}\right|\delta t \lambda_2\right)\right], \text{ for } n \neq n_0$$

$$P_{n_0} = Rr_{bg} + R(1-r_{bg})\left[1 - (1+a)\exp\left(-\frac{1}{4}\delta t \lambda_1\right) + a\exp\left(-\frac{1}{4}\delta t \lambda_2\right)\right],$$

(S.1)

where $\lambda_1 = \gamma_{EG} + \gamma_{GE}$, $\lambda_2 = \gamma_{MG} + \gamma_{EM}\gamma_{GE}/\lambda_1$ and $a = \lambda_2/\gamma_{MG} - 1$. The expression follows directly from the second-order autocorrelation function for a three-level system(*2*). The parameters of this model assumed in the simulation are summarized in Table S1.

### S2. Machine learning-based classifiers

Here we outline the details of the main binary classification algorithms used within this work: 1D CNN classifier, voting classifier, support vector classification based approach (SVC), and gradient boosting classifier. 1D CNN classifier was implemented using *Keras open-source machine learning* library(*4*), while the latter three were realized with the *Scikit-learn package for Python*(*5*).

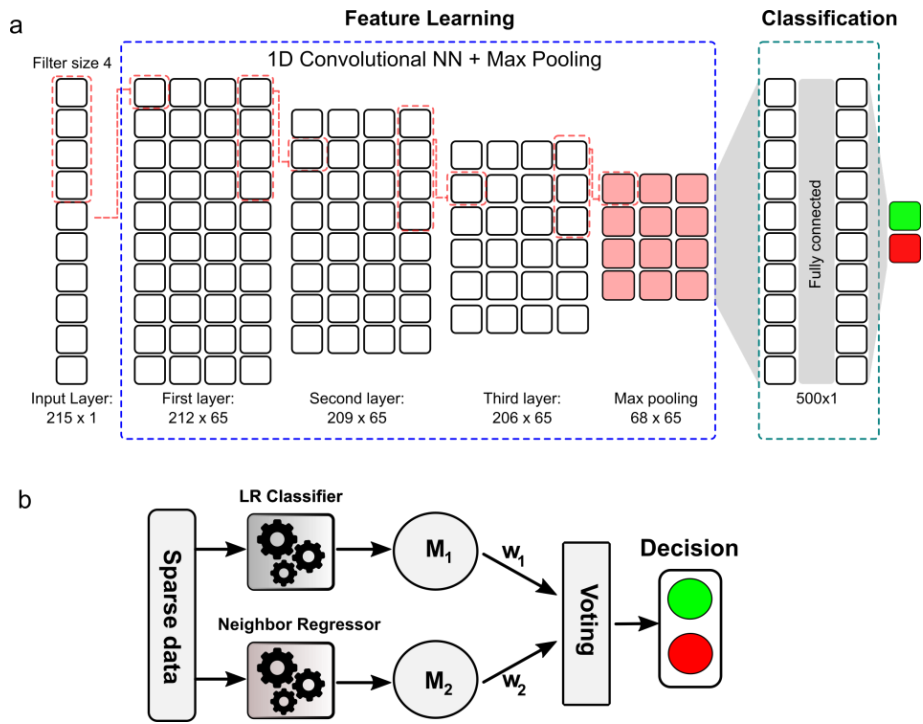

**Figure S1.** (a) Schematics of the 1D CNN classifier, (b) schematics of VC classification algorithm

- **1D convolutional neural network:** 1D CNN binary classifier consisted of one input layer, three hidden convolutional layers, one max-pooling layer followed by two fully connected layers and two output nodes containing the classification result. The input layer had 215 nodes corresponding to the number of bins in the autocorrelation experiment. The filters distribution in each hidden layer was optimized to capture the salient features of the autocorrelation datasets. All of the the hidden layer comprised 65 filters. The third layer was connected with the max-pooling layer. The kernel size of the filters(4) was chosen to be the same for each layer. The two fully connected layers performed binary classification and reduced the vector with a height of 500 to a binary vector. 80% of the generated datasets were used for used for CNN training while the remaining 20% were used for validation. Stochastic descent gradient optimizer was used for weight optimization with a learning rate of 0.01, a momentum of 0.9, and a decay rate of $10^{-5}$. The batch size was 150. The schematics of the 1D CNN classifier is shown in Figure S1a.

- **Voting classifier (VC):** The main idea of VC method is to combine fundamentally different supervised classification techniques by voting. In this work, we used soft voting by averaging the output probabilities produced by each classifier. Within this work, we applied a VC that used logistic regression (LG) and K-nearest neighbors (k-NN) methods. Specific weight distribution was assigned for both classifiers ($w_1 = 2$ for LG and $w_2 = 1$ for k-NN classifiers). The predicted class probabilities ($M_1$ and $M_2$) multiplied by corresponding weights, and averaged. The final voting is made by choosing the highest resulting averaged probability. The LG is set to use limited memory BFGS (Broyden–Fletcher–Goldfarb–Shannon method) optimization algorithm(*6*), while the k-NN uses 3 neighbors with uniformly distributed weights. The schematics of the VC classifier is shown in Figure S1b.

For the sake of the comparison along with 1D CNN and VC approaches, we have implemented support vector classification (SVC) and gradient boosting algorithm (G.Boosting)(*7*).

- **SVC** is a supervised machine learning technique that performs classification by identifying the partition of an *n*-dimensional space by hyperplanes that differentiates the classes with the highest accuracy. We used the radial-basis function (RBF) kernel

$$K(x, x') = \exp\left(-\gamma \|x - x'\|^2\right) \tag{S.2}$$

where $\|x - x'\|$ is the Euclidean distance between two points in a dataset, $\gamma$ is a parameter of the RBF kernel which equal to $\gamma = 1/n_{feature}$ ($n_{feature}$ is a number of main features in the training set).

- **Gradient boosting** method builds an additive model in a forward stage-wise fashion; it allows for the optimization of arbitrary differentiable loss functions. Here we used GB with learning rate = 0.1 with 1000 boosting stages.

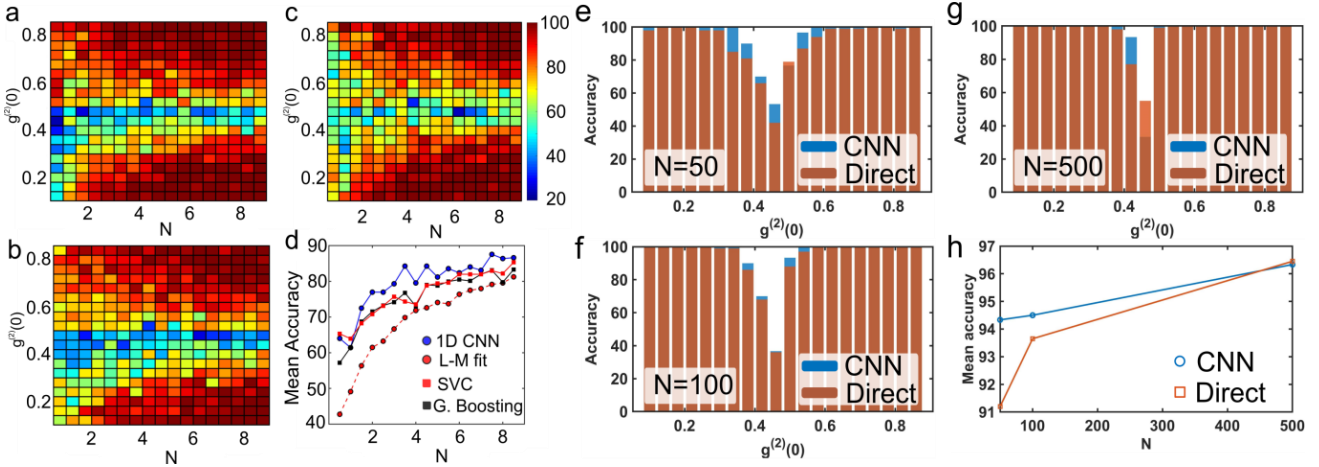

**Figure S2.** Accuracy distribution as a function of $g^{(2)}(0)$ and $N$ for: (a) gradient boosting classifier; (b) 1D CNN binary classifier and (c) SVC approach. (d) Averaged accuracy over all $g^{(2)}(0)$ realizations as a function of luminescence level $N$: CNN classifier (blue circles), SVC (red squares), gradient boosting (black squares), L-M fitting (red circles). (e)-(g) Accuracy as a function of $g^{(2)}(0)$ for three different cases of $N$ values: $N=50$ (e), $N=100$ (f) and $N=500$ (g). (h) Averaged accuracy over all $g^{(2)}(0)$ realizations as a function of large $N$ for CNN method and L-M fitting

A comparison of the mean accuracy of SVC and gradient boosting algorithms vs 1D-CNN is shown in Figure S2. Figures S2a-c show the accuracy distribution as a function of $g^{(2)}(0)$ and $N$ for: (a) gradient boosting classifier; (b) 1D CNN binary classifier and (c) SVC approach. Here we can see that all three ML-based approaches show similar performance. However, analysis of the averaged accuracy overall $g^{(2)}(0)$ shown in Figure S2d reviles that the CNN classifier shows slightly better accuracy.

**Classification of datasets with large average co-detection counts per bin.** Yet another open question is the performance of ML assisted classification algorithms when applied to datasets with large $N$. To be able to analyse this, we have performed the classification of simulated datasets with $N=50$, $N=100$ and $N=500$. Each of the considered cases contained 100 datasets. The training of the CNN classifier was done using the training set used in the main text combined with three aforementioned cases. 70% of the dataset was used for training, while the rest 30% used for testing. Figure S2e-g shows the accuracy distribution comparison for the three different cases $N$. As we can see here for large $N$ values, the accuracy trough becomes expectedly narrower, and classification methods perform almost identically. Figure S2h shows the dependence of the mean accuracy of the three methods as a function of $N$. Here we can see that indeed, with increasing $N$ values all three methods asymptotically tend to 100% accuracy.

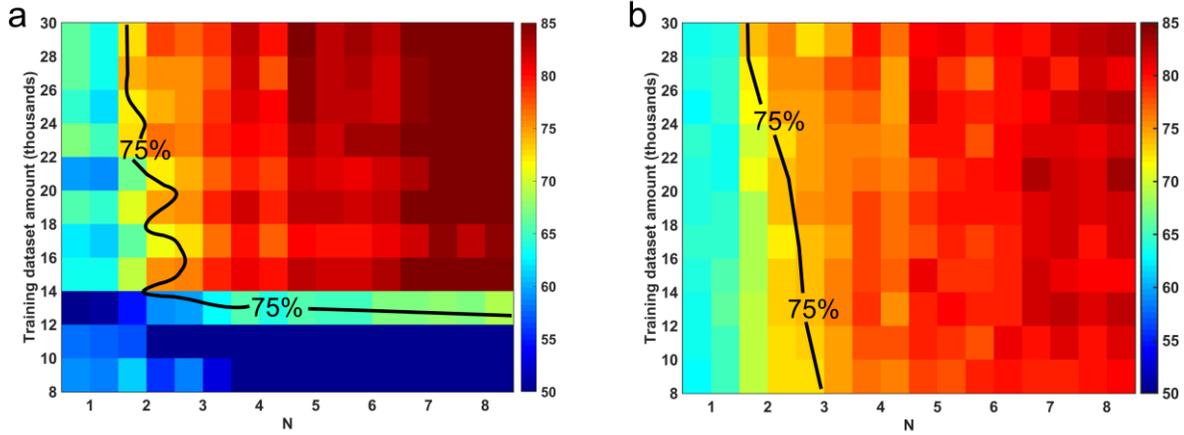

**Figure S3. Accuracy distribution averaged over $g^{(2)}(0)$ as a function of training dataset amount** for (a) CNN classifier (blue circles) and (b) VC approach. Contour line shows the boundary of >75% accuracy region

## S3. Dataset size influence on 1D-CNN and VC training:

The experimental autocorrelation data is more costly than the data obtained from the numerical experiment. Therefore, the training dataset for the neural network performing classification of physical emitters is limited. In order to train our networks on experimental data, we'd like to study how the classification accuracy depends on the size of the training dataset. We have tested CNN and VC classifier with different amounts of training dataset retrieved from the ETCE model. The training dataset is varied from 20% to 75% of total ETCE model data, i.e. changed from 8000 to 30000 datasets. Figure S3 shows the averaged accuracy over $g^{(2)}(0)$ as a function of $N$ and training dataset amount. The figure demonstrates that CNN classifier requires at least 14000 datasets to achieve a good classification score, while the VC approach shows good performance for all training dataset sizes.

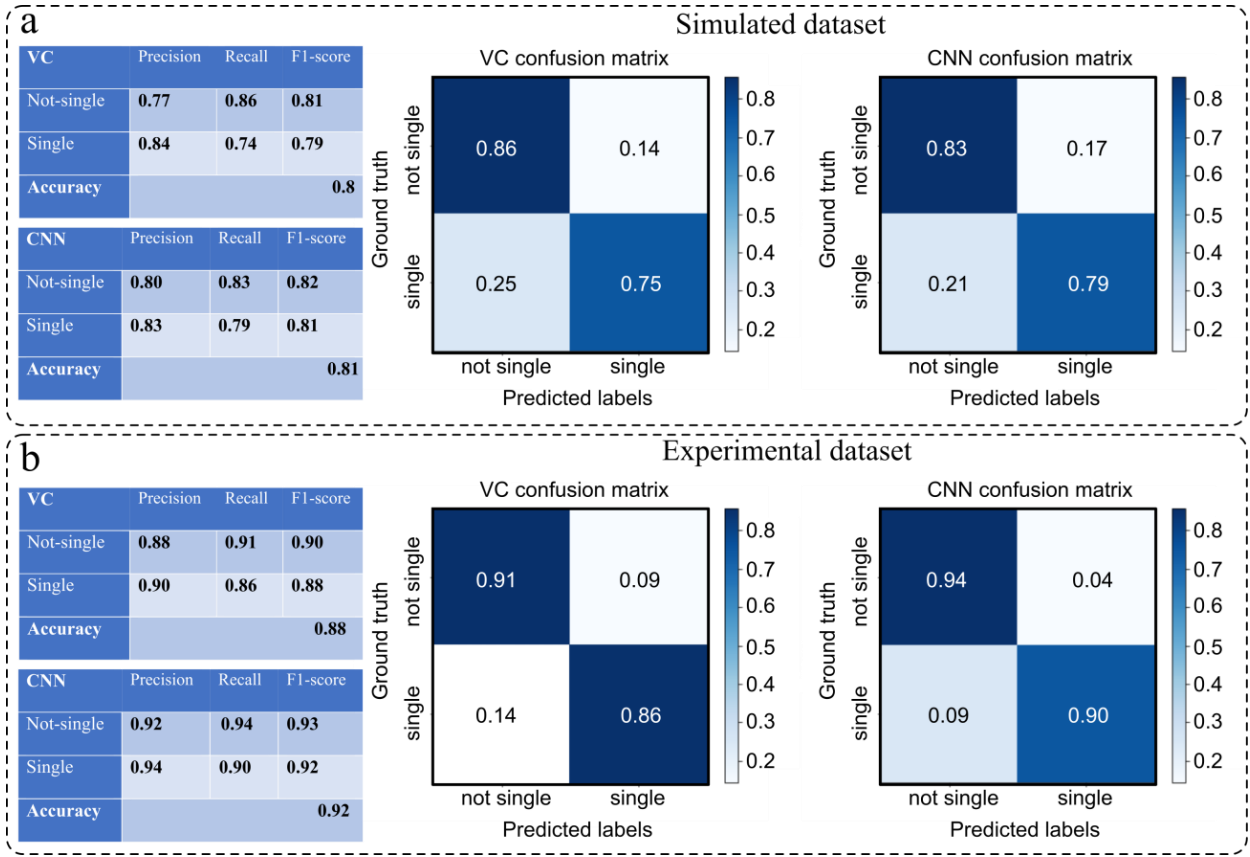

**Figure S4. Main performance specs of the voiting and CNN based classifiers**: precision, accuracy, recall, F-score and corresponding confusion matrices for simulated data (a) and experimental datasets (b)

## S4. Classifier performance

To be able to justify the performance of the proposed classification algorithms, along with the accuracy scores shown in the main text, we have calculated additional classification metrics of CNN-based classifier and VC algorithm. Particularly we have estimated precision, recall and F1-score of classifiers trained on simulated and experimental datasets (Fig. S4). Based on this analysis, we can see that both of the classifiers along with the high mean accuracy scores, ensure a balanced performance of the classification. To further clarify this statement, Fig. S4 shows the confusion matrices for both classification schemes. Both methods exhibit balanced performance on distinguishing between "single" and "non-single" emitters. In both cases, the CNN classifier outperforms VC classification results.

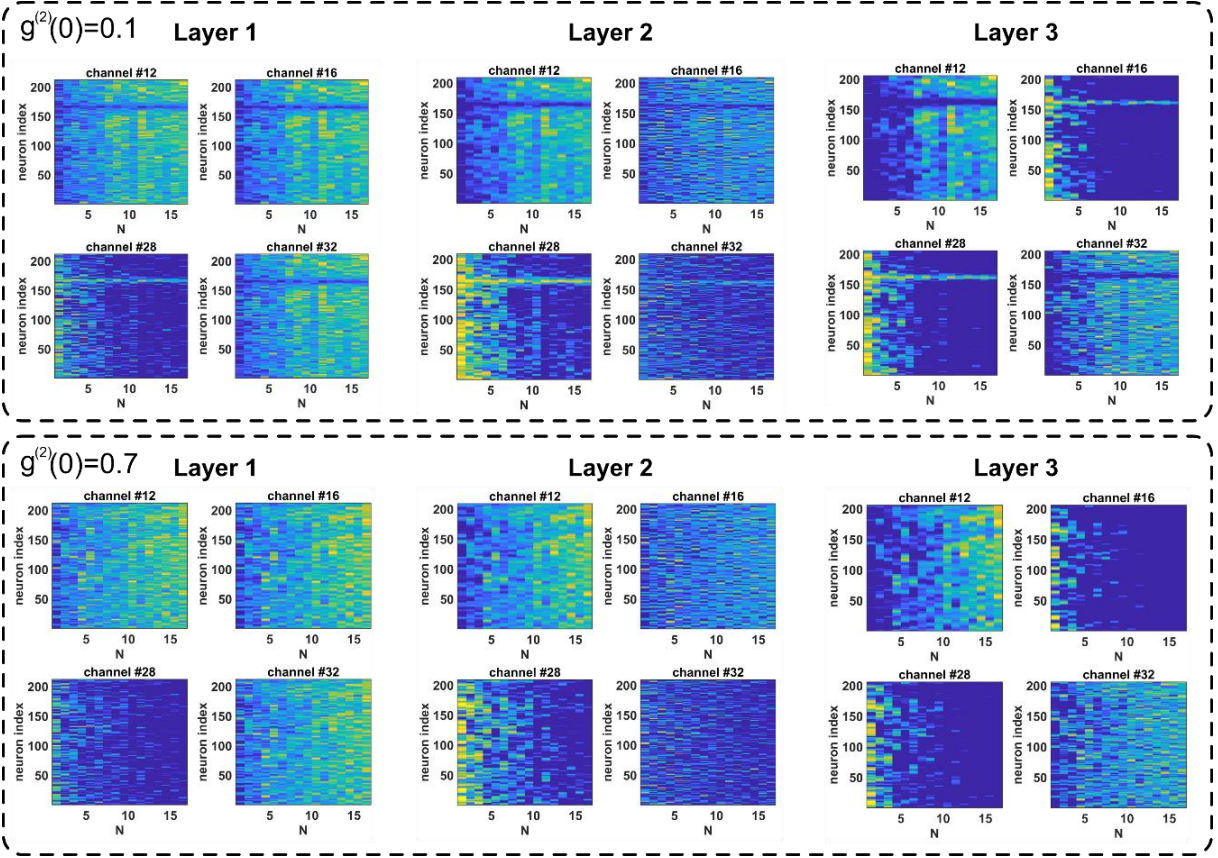

**Figure S5. Feature map of the convolutional layers.**

**S5. Class activation map of CNN classifier** Yet another important question that should be clarified is the decision-making mechanism of the CNN network. To be able to gain more intuition on the CNN performance we have determined the feature maps of the trained CNN classifier within each convolutional layer. Particularly we have passed two test samples corresponding to two classes through the network and captured the outputs of each layer. This allows us to understand what features are learned on each layer. Figure S5 shows the feature maps as a function of neuron index and average counts per bin values for two different datasets: "single" class ($g^{(2)}(0) = 0.1$) and "not-single" class ($g^{(2)}(0) = 0.7$). Here we are showing only 4 channels out of 65 for each layer. Here we can see two main facts:

(i) **Different channels are learning different areas of average counts per bin range**. Here we can see that some channels activate only for sparse datasets (e.g. channels 16 and 28 of layer 3), while some channels focus on the datasets with higher N values (e.g. channels 12 and 32 of layer 3);

(ii) **The main features of both classes are quite different**. Here we can clearly see that for "single quantum emitter" class, the region of the zeroth time bin is in the focus of all channels. At the same time for "not single quantum emitter" class, the features are spread around all of the neurons.

To be able to gain more insight into the CNN performance, we have implemented a gradient-weighted class activation mapping (grad-CAM) algorithm (*8*), which allows visualizing of neuron activation patterns inside the last convolutional layer in different test dataset cases. This analysis allows determining what features of the sparse autocorrelation map play the most important role in the decision making of the CNN based classifier. Particularly, the grad-CAM exploits the gradient information flowing into the last convolutional layer to understand the role of each neuron on the decision-making process. In order to do it, we have calculated the gradient of the score for a given class over the convolutional layer's feature maps $A^k$, i.e. $\partial y^{class} / \partial A^k$. This allows us to determine the important weights by means of the gradient average pooling:

$$\alpha_k^{class} = \frac{1}{Z} \sum_i \sum_j \frac{\partial y^{class}}{\partial A^k} \quad (S.4)$$

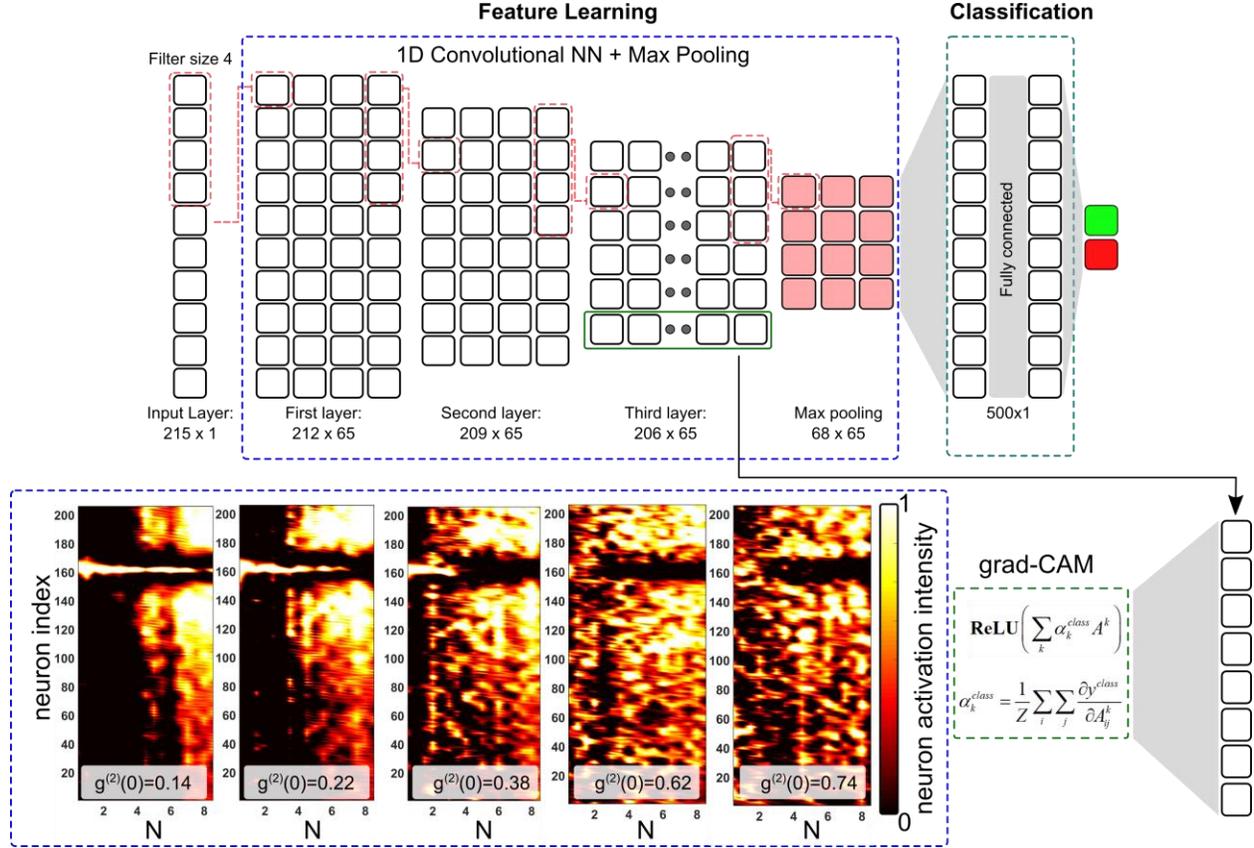

**Figure S6. Schematics of grad-CAM algorithm.** The lower section shows the dependences of class activation maps on $N$ values for 5 different cases of $g^{(2)}(0)$: 0.14, 0.22, 0.38, 0.62 and 0.74.

Finally, the class discriminative localization map $L^{class}$ is determined by applying the ReLU activation function on the weighted combination of forward activation maps:

$$L^{class} = \mathbf{ReLU}\left(\sum_k \alpha_k^{class} A^k\right) \quad (S.5)$$

In the case under consideration, the grad-CAN analysis leads to 206-dimensional class activation vector due to the dimension of the third convolutional layer. We applied this technique to the simulated dataset corresponding to 5 different cases of $g^{(2)}(0)$ 0.14, 0.22, 0.38, 0.62 and 0.74. Figure S6 shows schematics of the grad-CAM method, as well as class activation maps for 5 different cases of $g^{(2)}(0)$ plotted as a function of the average co-detection counts per bin $N$. This analysis reveals that activation maps of "single" and "not single" classes significantly different from each other and counterintuitively, not localized around the zero time delay bin. Particularly, in the case of a "single" emitter case, the CNN classifier looks mostly into the zeroth time delay bin area in the case of small values of $N$. With the increase of the average co-detection counts per bin value, the classifier is making the decision based on the two "wings" of the autocorrelation map around the zeroth time delay bin. More interestingly, for the case of "not single" quantum emitters, the CNN classifier makes the decision based on quite spread features of the autocorrelation maps. Here we can clearly see that along with the zero time delay bin information, the surrounding features play an important role for the decision-making mechanism of the CNN classifier.

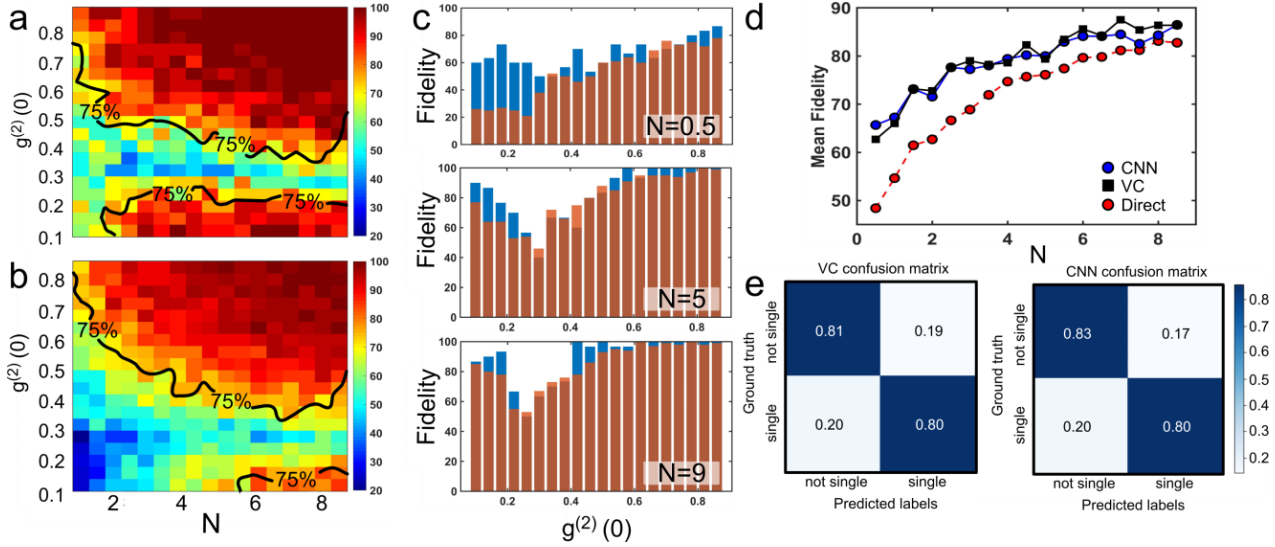

**Figure S7. Schematics of grad-CAM algorithm.** The lower section shows the dependences of class activation maps on $N$ values for 5 different cases of $g^{(2)}(0)$: 0.14, 0.22, 0.38, 0.62 and 0.74.

**S6. Classification of quantum emitters for decision criterion $g^{(2)}(0) = 0.3$**

In some applications, the purity of the emission could be more strictly constrained, which results in a lower level of the decision boundary between "single" and "not-single" classes. To be able to test the performance of the ML-based classification algorithms in the case of the shifted decision boundary, we have performed sparse data classification in the case of the following classification criteria: "single" $0 < g^{(2)}(0) \leq 0.3$ and "not-single" $0.3 < g^{(2)}(0)$. Based on the simulated dataset used in the main text we have retrained VC and CNN-based classifiers to distinguish between the aforementioned two classes. The accuracy map of CNN based classification and L-M fitting are shown in Fig.S7a-b, respectively. Due to the shifted classification criteria, as expected, the accuracy dip shifts to $g^{(2)}(0) = 0.3$ area. As in the case of "symmetric" classification considered in the main text, the CNN-based classifier shows superior performance in comparison with the L-M fitting, especially in the low $N$ area. The comparison between CNN and L-M fit for different cases of $N$ is shown on Fig.S7c. When the co-detection event number is low, L-M fitting results in unbalanced classification, while the CNN-based approach shows uniform performance. Fig. S7d shows the dependence of mean accuracy, averaged over all $g^{(2)}(0)$ values, on $N$. Based on this we can conclude that the CNN classifier and VC shows identical performance. Noting, that CNN classifier has not been optimized for the particular problem. We envision that with specific optimization of the CNN structure it is possible to increase the classification efficiency. The confusion matrices for both methods are shown on Fig. S7e. Here we can see that CNN and VC shows balanced classification of both classes.

**S7. Experimental data**

Physical autocorrelation measurements were performed on a set of 41 emitters. For each emitter, autocorrelation datasets were acquired in a series of 1-second intervals. The sparse datasets acquired for each emitter were compounded into a "full" dataset, from which the $g^{(2)}(0)$ value was attained using the L-M fitting algorithm. These fitted values of $g^{(2)}(0)$ on full datasets were appended to each corresponding sparse dataset as a label and used as the ground truth for the training/testing purposes. Detailed information about dataset for each emitter, along with the classification scores fidelities for VC and L-M fitting are presented in figure S8.

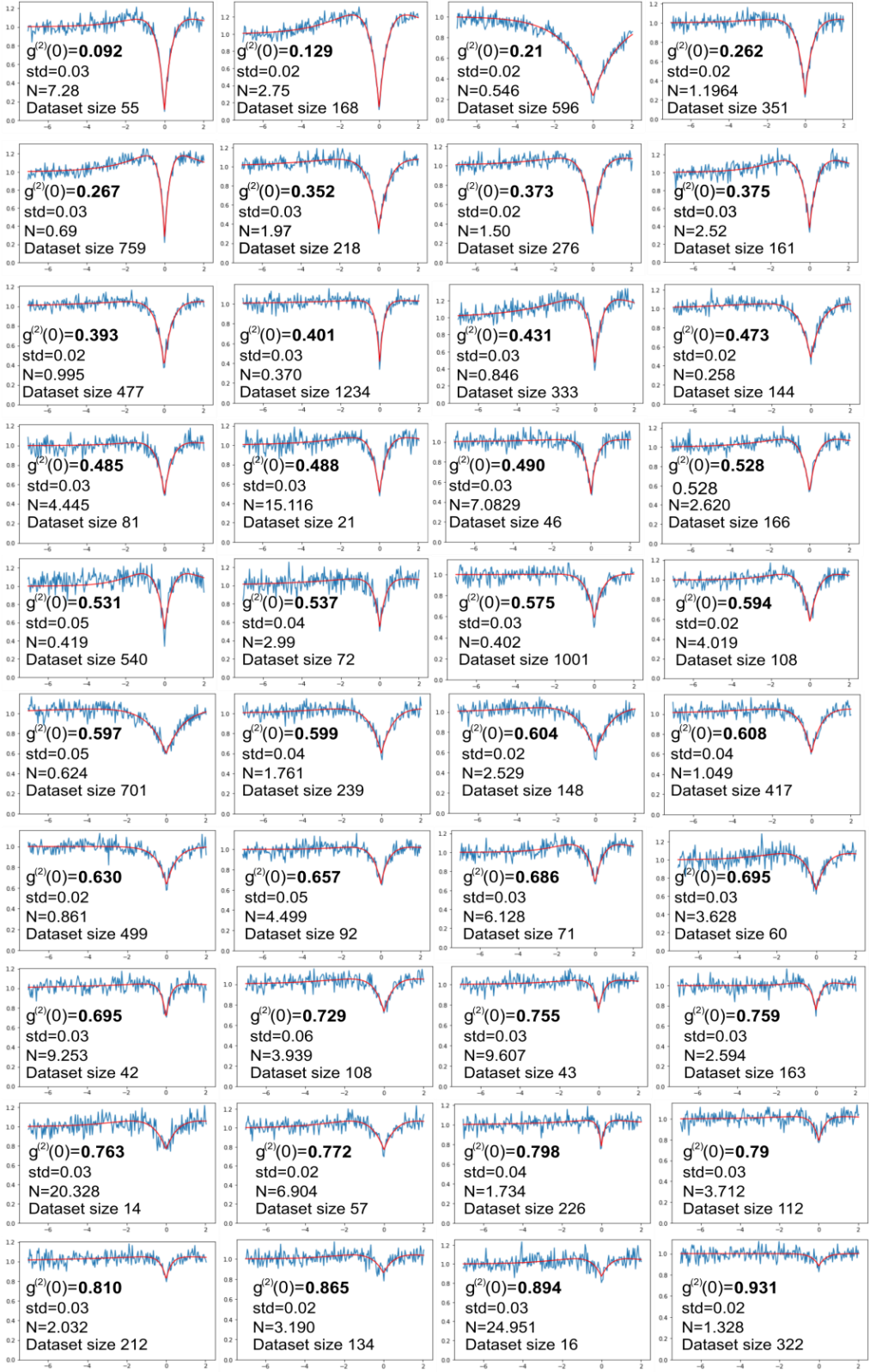

**Figure S8. Detailed information on datasets corresponding to each emitter collected from photon autocorrelation measurements.** The blue curves correspond to rough data, the red curves correspond to the LM fitting. The inset shows $g^{(2)}(0)$, $N$ values and the dataset size of the corresponding emitter, 'std' – stands for the standard deviation of the fit.

**Supplementary References**